\documentclass{pazha2} 
\usepackage{epsf} 
\usepackage{graphicx} 
\usepackage{multirow} 
\usepackage{pdflscape} 
\usepackage{color} 
\definecolor{darkblue}{rgb}{0,0,0.9} 
 
\def\*{$^{*}$}
\def\и{$^{\mbox{\small и}}$}
\def\к{$^{\mbox{\small к}}$}
\def\л{$^{\mbox{\small л}}$}
\def\ергс{эрг~с$^{-1}$}
\def\ергсм{эрг~см$^{-2}$~с$^{-1}$}

\def\aa{$^{\mbox{\small a}}$}
\def\bb{$^{\mbox{\small b}}$}
\def\cc{$^{\mbox{\small c}}$}
\def\dd{$^{\mbox{\small d}}$}
\def\ee{$^{\mbox{\small e}}$}
\def\ff{$^{\mbox{\small f}}$}
\def\gg{$^{\mbox{\small g}}$}
\def\hh{$^{\mbox{\small h}}$}
\def\ii{$^{\mbox{\small i}}$}

\begin{document}
\journalinfo{2019}{45}{11}{1}{768}{786}[16]
\sloppypar

\title{Observation of the second LIGO/Virgo event connected with
  binary neutron star merger S190425z in the gamma-ray range}
\year=2019 \author{
  A. S.~Pozanenko\address{1,2}\email{apozanen@iki.rssi.ru},
  P. Yu.~Minaev\address{1}, S. A.~Grebenev\address{1},
  I. V.~Chelovekov\address{1} \addresstext{1}{Space Research
    Institute, Russian Academy of Sciences, ul. Profsoyuznaya
    84/32, Moscow, 117997 Russia} \addresstext{2}{National
    Research University ``High School of Economics'',
    ul. Myasnitskaya 20, Москва, 101000 Russia}}

\shortauthor{POZANENKO et al.}
\shorttitle{OBSERVATION OF THE LIGO/VIRGO EVENT S190425Z IN THE GAMMA-RAY RANGE}  

\submitted{September 16, 2019}
\revised{October 10, 2019}
\accepted{October 22, 2019}

\begin{abstract}
\noindent
Observations of the gravitational-wave (GW) event S190425z
registered by the LIGO/Virgo detectors with the Anti-Coincidence
Shield (ACS) of the gamma-ray spectrometer SPI aboard the
INTEGRAL observatory are presented. With a high probability ($>
99$\%) it was associated with a neutron star merger in a close
binary system. This is only the second event of such type in the
history of gravitational-wave observations (after GW\,170817). A
weak gamma-ray burst, GRB\,190425, consisting of two pulses in
$\sim0.5$ and $\sim5.9$ s after the moment of merging the stars
in S190425z was detected by SPI-ACS. The pulses had a priori
reliability of $3.5\sigma\ \mbox{and}\ 4.4\ \sigma$ as single
events and $5.5\sigma$ as a combined event. Analysis of the
SPI-ACS count rate history recorded these days ($\sim 125$ ks
observations in total) has shown that the rate of the appearance
of two close pulses with characteristics of GRB\,190425 by
chance does not exceed $6.4\times 10^{-5}\ \mbox {s}^{-1}$ (that
is, by chance such events occur on average every $\sim4.3$
hours). We note that the time profile of the gamma-ray burst
GRB\,190425 has a lot in common with the profile of the
gamma-ray burst GRB\,170817A accompanying the GW\,170817 event;
that both the mergers of neutron stars were the closest
($\la150$ Mpc) of all the events registered by the LIGO/Virgo
detectors; and that there were no confident excesses of
gamma-ray emission over the background detected in any of $\sim
30$ black hole merger events recorded to the moment by these
detectors. No hard X-ray flares were detected in the field of
view of the SPI and IBIS-ISGRI gamma-ray telescopes aboard
INTEGRAL. This, as well as the lack of detection of gamma-ray
emission from GRB\,190425 by the GBM gamma-ray burst monitor of
the {\sl Fermi\,} observatory assuming its occultation by the
Earth, can significantly reduce the localization area for the
source of this gravitational-wave event. The estimates of the
parameters $E_{\rm iso}$ and $E_{\rm p}$ for the gamma-ray burst
GRB\,190425 are obtained and compared with the similar
parameters for the GRB\,170817A burst.\\

\noindent
    {\bf DOI:} 10.1134/S032001081911007X\\
    
\end{abstract}

\section{INTRODUCTION}
\noindent
Registration of the gravitational-wave signal GW\,150914 from
the merger of two black holes (Abbott et al. 2016) marked the
beginning of the era of gravitational wave astronomy. Over the
past four years the Advanced LIGO gravitational-wave detectors
and the Advanced Virgo detector, that started to operate in
August 2017, (hereafter just LIGO and Virgo) have already
recorded $\sim40$ similar events. Sensitivity of the detectors
is growing rapidly: there were only 3 events recorded in the O1
cycle of the LIGO work (since September 12, 2015 till January
19, 2016), 8 events recorded in the O2 cycle (since November 23,
2016 till August 25, 2017), and already 31 events recorded in O3
(started on April 1, 2019) by the end of September 2019. The
catalog of events for cycles O1--O2 can be found in Abbott et
al. (2019), the current list of O3 events --- on the web-site
{\sl gracedb.ligo.org/superevents/public/O3\/}.

The LIGO/Virgo detectors are optimized for observing signals
from compact binaries, therefore, they can successfully register
not only mergers of binary black holes (BBH), but also the
systems of black hole --- neutron star (NSBH) or binary neutron
stars (BNS). The frequency of recording various types of events
depends on the viewed volume of the local Universe and the
number of systems of a certain type in this volume. The volume
itself is proportional to the 3rd power of the distance from
which it is possible to register a signal with a minimum
amplitude, and the distance --- to the mass of the lightest
component of a binary system. No wonder that the amount of the
registered BBH mergers is far outnumber the mergers of NSBH and
BNS --- at the time of submitting the paper there were 2
reliable (having a probability of $\ga85$\%) signals recorded
from NSBH (S190814bv and S190910d) and 3 --- from BNS
(GW\,170817, S190425z, S190901ap), all of these systems were
located at significantly shorter distances than the recorded BBH
mergers.

Intensive search for bursts of electromagnetic radiation during
and after each LIGO/Virgo event led to the only reliable
detection --- the gamma-ray burst GRB\,170817A (Goldstein et
al. 2017; Savchenko et al. 2017c; Pozanenko et al. 2018),
accompanying the first event, GW,170817, discovered from BNS
mergers (Abbott et al. 2017a, 2017b). This is fully consistent
with theoretical expectations --- no effective mechanism for
the electromagnetic pulse formation during the BBH merging has
been proposed to the moment, and the probability of recording
the emission from NSBH events is rated very low (e.g. Postnov et
al. 2019). The gamma-ray burst GRB\,170817A was observed with a
delay of $\sim 1.7$ s relative to the registration time $T_0$ of
the gravitational-wave event, that is, the gamma-ray emission
was formed already after the merger of binary neutron stars. It
also meets expectations. There was the kilonova AT2017gfo
detected in the direction of the burst arrival, in the galaxy
NGC\,4993 (Coulter et al. 2017; Evans et al. 2017; Troja et
al. 2017). Its observations allowed this unusual type of
supernovae to be studied in detail for the first time.

Search for electromagnetic emission from gravitational-wave
events is carried out primarily in the hard X-ray or soft
gamma-ray ranges, as well as in optics. Appearance of the hard
emission in the form of a short gamma-ray burst due to the merge
of BNS (and NSBH) systems has been predicted by Blinnikov et
al. (1984) and Paczynski (1991). Modern all-sky monitors are
capable to successfully detect such bursts in hard X-rays even
at distances of dozens of Gpc. The burst detection is of extreme
importance because it allows the localization region of the
gravitational-wave event determined by triangulation 
of signals measured by the LIGO detectors L1 and L2, located in
the USA, and the Virgo detector V1, located in Italy, to be
notably reduced. In the O3 cycle of operation of the
gravitational antennas the minimum localization area for an
event was 23 squared degrees, and the maximum area was more than
24 thousand squared degrees.

In optics, the transient associated with the appearance of a
kilonova or a gamma-ray burst afterglow is expected although the
direct search for optical transients in such large localization
areas is extremely challenging. Nevertheless, many observatories
and network projects are involved in solving the problem. Two
tactics are used: (1) wide-angle telescopes carry out mosaic
scanning of the entire area of localization, (2) narrow-angle
ones are successively observing galaxies located in the
three-dimensional volume of localization, determined by the
solid angle and the range of possible distances to the
source. The number of galaxies in such a volume can reach tens
of thousands, nevertheless, their successive viewing turns out
to be more effective than scanning of the entire
area. Unfortunately, the existing catalogs of galaxies are not
complete, and it is not always possible to confine yourself by
only to the second tactic when searching for an optical
component of the gravitational wave event.

Taking into account the importance of timely detection of
gamma-ray bursts concomitant to the merger of BNS or NSBH, we at
the Space Research Institute of the Russian Academy of Sciences
(IKI RAS) have initiated the works under the program of
searching for a transient hard X-ray emission from all such
events recorded by the LIGO and Virgo detectors. The open-access
data from the SPI-ACS and IBIS-ISGRI gamma-ray telescopes of the
INTEGRAL astrophysical observatory were used for the search. In
the case of S190425z (the second recorded merger of BNS) such
emission has been found (Minaev et al. 2019; Chelovekov et
al. 2019a; see also Martin-Carillo et al. 2019; Savchenko et
al. 2019).

In the present paper, we describe in detail the results of these
observations, compare the found gamma-ray burst with the
GRB\,170817A burst accompanying the first event of a BNS merger
(GW\,170817), and consider all available arguments in favor of the
reliability of its detection.

\section*{INSTRUMENTS AND METHODS}
\noindent
As already mentioned, this study is based on observations of two
main instruments aboard the international gamma-ray astrophysics
observatory INTEGRAL (Winkler et al. 2003): the gamma-ray
telescope IBIS-ISGRI (Lebrun et al. 2003; Ubertini et al.  2003)
and the gamma-ray spectrometer SPI (Vedrenne et al. 2003; Rock
et al. 2003). For sky imaging and individual cosmic source
studying the principle of coded aperture is used in both
telescopes.

The IBIS gamma-ray telescope is designed for mapping the sky in
hard X-ray and soft gamma-ray ranges and studing detected
sources with rather a rough energy resolution ($E/\Delta E \sim
13$ at 100 keV). The telescope has a field of view (FWZR) of
$30\deg \times 30\deg$ in size and angular resolution of
12\arcmin\ (FWHM). The position of the bright bursts may be
determined with precision of $\la 2$\arcmin. Maximum sensitivity
of the ISGRI detector of this telescope, which is an array of
16384 CdTe elements, falls in the 18--200 keV range. Its total
area is 2620 cm$^2$, effective area for the events in the center
of the field of view $\sim 1100$cm$^2$ (half is obscured by
opaque mask elements).

The SPI gamma-ray spectrometer is designed for thin ($E/\Delta
E\sim 550$ at 1.7 MeV) gamma-ray spectroscopy of cosmic
annihilation radiation (from the central regions of the Galaxy)
and radiation in nuclear gamma-ray lines of radioactive origin
(from remnants of young nearby supernovae). The telescope has 
maximum sensitivity in the range 0.05--8 MeV, a hexagonal field
of view with a diameter of $32\deg$ (FWZR) and angular resolution
$2\fdg5$ (FWHM); the geometrical area of 19 cryogenic superpure
Ge detectors is $\simeq 500$ cm$^2$.

For timely detection and identification of gamma-ray bursts and
other transient events caught in the field of view of the IBIS
and SPI telescopes, as well as for urgent notification of them
via electronic GCN (Gamma-ray Coordinates Network) circulars the
IBAS automatic software system was developed (Mereghetti et
al. 2003) and is now successfully used. Yet there is always a
possibility to independently search for bursts in just received
or even archival data of the telescopes. In such a way some
gamma-ray bursts, that were not recorded by the IBAS system for
different reasons, were found (Grebenev, Chelovekov 2007;
Minaev et al. 2012, 2014; Chelovekov et al. 2019b). Analysis of
such data can be carried out using the standard software package
for the INTEGRAL observatory -- OSA.  In this work, we have used
the OSA version 10.2 package.

\subsection*{ACS shield of the SPI gamma-ray spectrometer}
\noindent
Although some gamma-ray bursts are successfully detected by the
SPI gamma-ray spectrometer and IBIS-ISGRI gamma-ray telescope
within their fields of view (e.g., Mereghetti et al. 2003; Foley
et al. 2008, 2009; Vianello et al. 2009; Minaev et al., 2014)
and beyond (e.g., Minaev et al. 2014; Chelovekov et al. 2019b),
significantly more of them are registered by anti-coincidence
shield ACS of the SPI spectrometer, which has a much larger area
(Rau et al. 2004, 2005)\footnote{The IBIS-ISGRI detector is also
  equipped with an active anti-coincidence shield, the so-called
  VETO System, but its data are grouped and transmitted to the
  Earth in 8-s intervals that makes it unsuitable for searching
  for short gamma-ray bursts (Ubertini et al. 2003; Quadrini et
  al. 2003).} The SPI-ACS shield is one of the most sensitive
all-sky monitors in the history of gamma-ray burst
observations. Thanks to the highly elliptical orbit of INTEGRAL
(Eismont et al. 2003) with a period of 72 hours (64.8 hours
after 2015), there are almost no zones for it shaded by the
Earth (coverage is $\ga80$\% of the sky), while a stable
background on a time scale of hundreds or even thousands of
seconds allows it to effectively carry out a sub-threshold
search for transients of different durations. SPI-ACS has been
successfully used in searching for transient gamma-ray emission
from the gravitational-wave event GW\,170817 (Savchenko et
al. 2017c; Pozanenko et al. 2018), having confirmed the
detection of the GRB\,170817A gamma-ray burst by the {\sl
  Fermi\,}/GBM monitor (Goldstein et al.  2017). It was again the
SPI-ACS data that we begin the present study with.

The SPI-ACS shield consists of 91 scintillation crystals Bi$
_4$Ge$_3$O$_{12}$ (BGO) with a total mass of 512 kg (von Kienlin
et al. 2003a, 2003b; Ryde et al. 2003). Their total effective
area for detection of gamma-ray bursts reaches
$\sim0.7\ \mbox{m}^2$. The total count rate is transmitted to
the Earth with a resolution of 50 ms, telemetry contains no
spatial or spectral information. The energy range is known
poorly, because the parameters of individual photomultipliers
and light outputs of the crystals vary slighly and are known
inaccurately. The lower threshold can be roughly estimated to be
$\sim 80$ keV, the upper one --- $\ga10 $ MeV. Due to some
specifics of the geometrical construction of SPI, its shield,
being almost omnidirectional, is insensitive to the bursts coming at
small angles to axis of the telescope.

\subsection*{Method of searching for gamma-ray bursts}
\noindent
Data of the SPI-ACS detector are a record of photon count rate
in the single broad energy band, however, the approximation of the
average count rate, estimation and subtraction of the
background, search for gamma-ray bursts from these data and
analysis of their reliability can be carried out using different
techniques (see, for example, Mereghetti et al. 2003; Savchenko
et al. 2012, 2017a; Minaev et al. 2014; Minaev, Pozanenko,
2017). Using different techniques, it is possible to obtain
slightly different results. As we will see, the important factor
affecting the results is the choice of an adequate time scale for
the analysis (size of the time bin in the studied light curve).

When searching for short transient flashes in the SPI-ACS data,
connected with LIGO/Virgo gravitational-wave events, we used the
following technique. First, at time intervals
($T_0-200\ \mbox{s}$, $T_0-50\ \mbox{s}$) and
($T_0+50\ \mbox{s}$, $T_0+200\ \mbox {s}$) equally spaced from
the moment $T_0$ of the arrival of the gravitational-wave
signal, the count rate was approximated by polynomial models of
the 1st and 3rd order using SPI-ACS data of initial resolution
(50 ms). The best residual model was taken as a background
model. The sample variance of the count rate history was
computed relative to this model. Note that the sample variance
of the SPI-ACS data differs from the Poissonian one by a factor
of 1.2--1.6 times (von Kienlin et al. 2003a; Ryde et al. 2003;
Rau et al. 2004, 2005). The adopted background model was
extrapolated to the time interval ($T_0-30\ \mbox{s}$,
$T_0+30\ \mbox{s}$) where our search for significant excesses in
the count rate over background was carried out. Time series with
different lengths of step (bin), from 0.1 to 10 s (of course, a
multiple of the bin length of the original time series 50 ms),
have been used.  Significance of the excess detected in some bin
of the count rate history over the background rate was estimated
using the sample variance reduced to the selected bin size. The
algorithm is optimal for searching for a pulse signal with
duration approximately matching the bin length selected for the
search (from 0.1 to 10 s).

\begin{figure}[t]
  \includegraphics[width=0.99\linewidth]{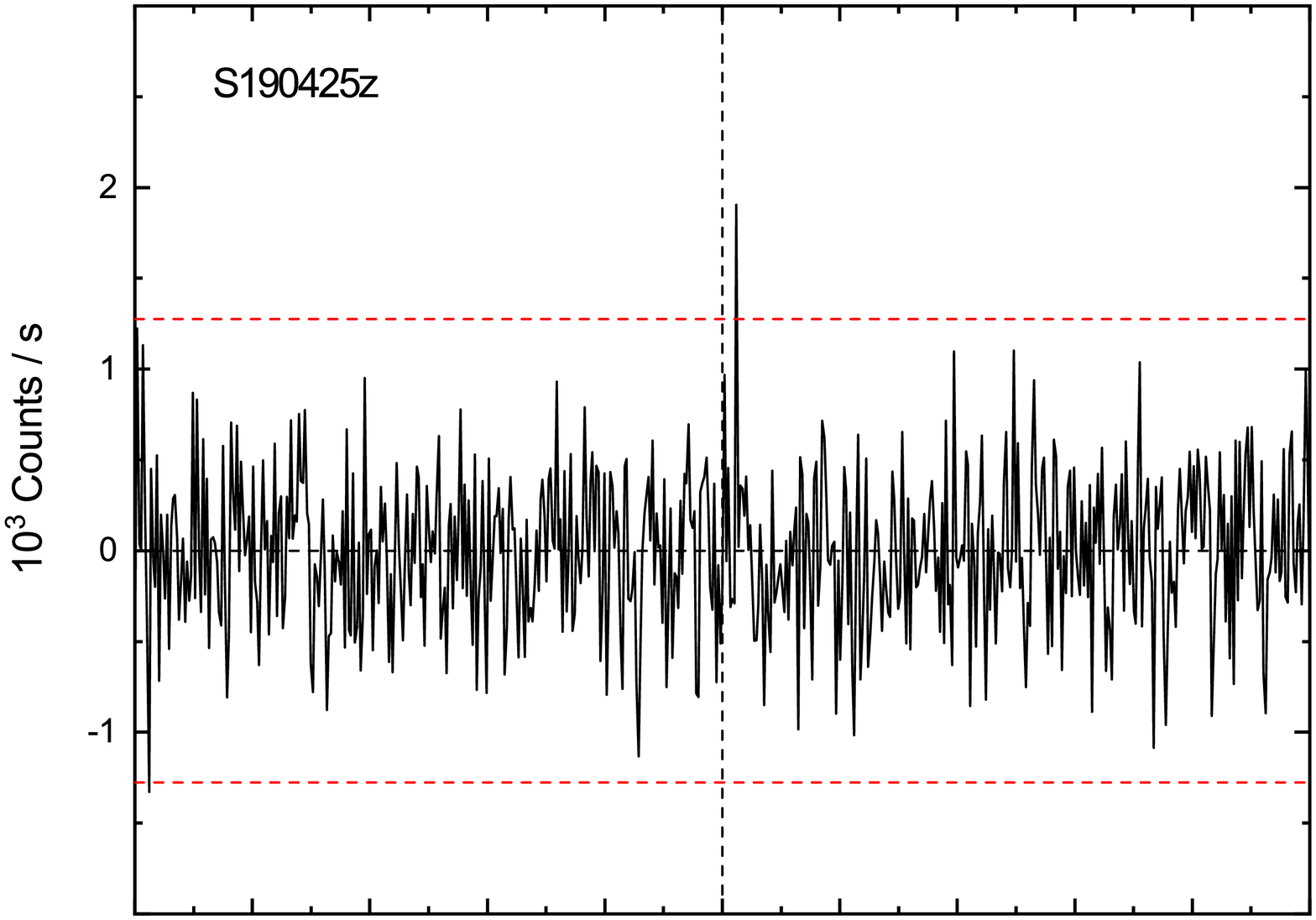}
  \includegraphics[width=0.99\linewidth]{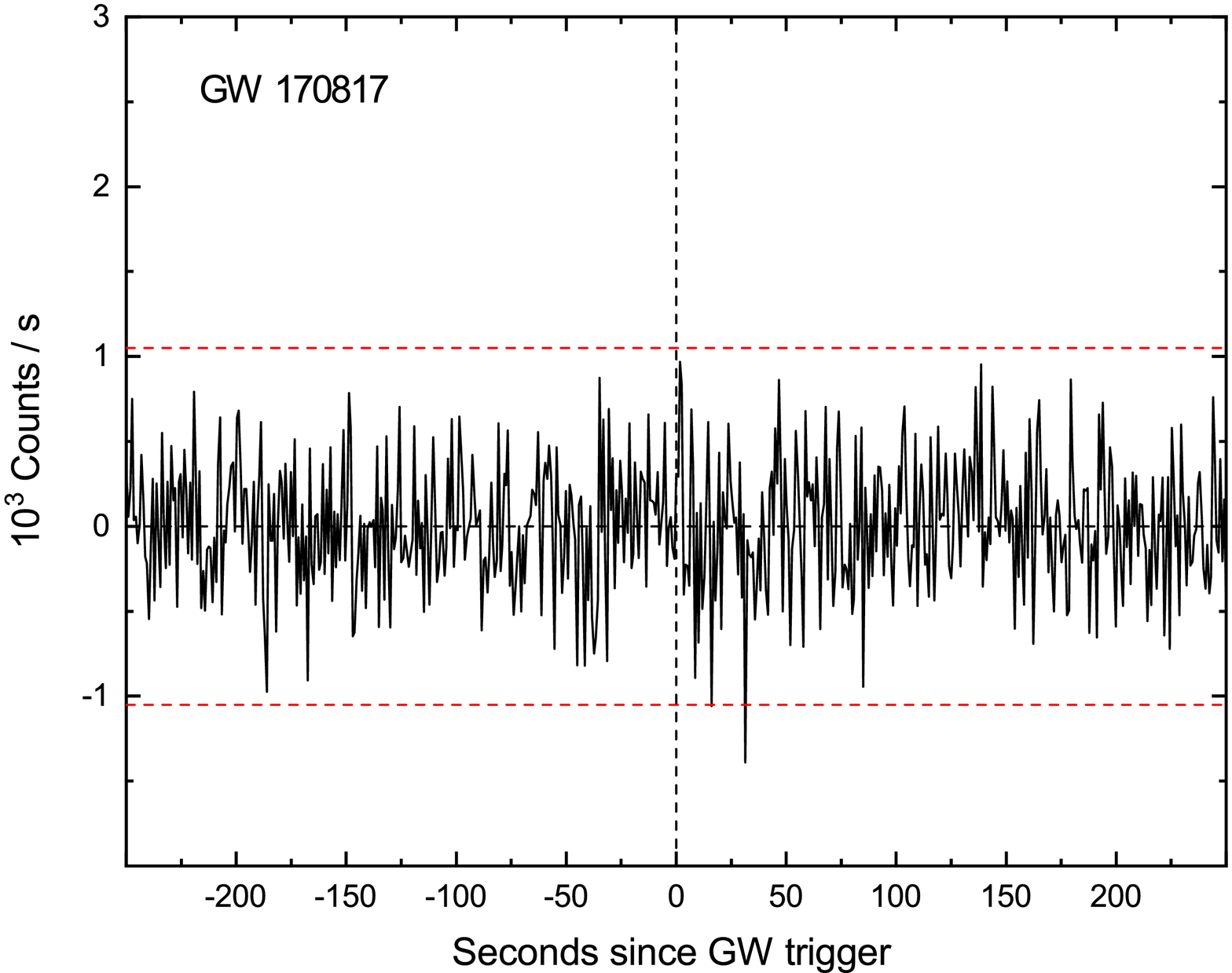}
\caption{\rm Photon count rate by the SPI-ACS detector versus
  time immediately before and after ($\pm250 $ s) the
  gravitational-wave events S190425z (top) and GW\,170817
  (bottom). The bin duration is 0.85~s, time is counted from the
  moment of recording the events by the LIGO/ Virgo detectors
  (vertical dashed line). The background is subtracted according
  to the model. Dashed (red) lines indicate the range of random
  deviations at the level of $3\sigma$.
\label{fig:lcurve_0.85}}
\end{figure}

The search for short gamma-ray bursts in data of the SPI
spectrometer is described in detail by Minaev et al. (2014), the
search for bursts in data of the IBIS-ISGRI telescope --- in the
work of Chelovekov et al. (2019b). In our paper this technique
(differing only by the choice of a smaller step for the analyzed
time series) is used for the search for a burst of hard X-ray
and soft gamma-ray emission associated with a gravitational-wave
signal. The same set of steps was taken that was used to
construct light curves from data of the SPI-ACS detector.

\begin{table*}[t]
  
\noindent
{\bf Table 1.} The main parameters of gravitational-wave events
GW\,170817 and S190425z and the gamma-ray bursts accompanying
them GRB\,170817A and GRB\,190425.\\ [2mm]

\begin{tabular}{l|c|c@{}c|c}\hline\hline
  &&\multicolumn{2}{|c}{}&\\ [-3mm]
LIGO/Virgo event&S190425z&&\multicolumn{2}{c}{GW\,170817}\\  \hline
&&&\multicolumn{2}{c}{}\\ [-3mm]
Trigger time $T_0$\aa& 2019-04-25 08:18:05&\makebox[5mm]{}
&\multicolumn{2}{c}{2017-08-17 12:41:04}\\ 
Distance to the source, Mpc& $156\pm41$&&\multicolumn{2}{c}{$40\pm8$}\\
Localization area (90\%)\bb, sq. deg.  &7461&&\multicolumn{2}{c}{16}\\
Angle to the axis of the SPI-ACS detector& 26\deg -- 60\deg&&\multicolumn{2}{c}{~105\deg}\\ \hline
 &&\multicolumn{2}{|c}{}&\\ [-3mm]
Gamma-ray burst&GRB\,190425&&\multicolumn{2}{c}{GRB\,170817A}\\ \hline
&&&&\\ [-3mm]
Pulse in the GRB profile
&first+second&\multicolumn{2}{c|}{first}&first+second\\ 
&&&&\\ [-3mm]
Experiment&  SPI-ACS &\multicolumn{2}{c|}{SPI-ACS} &
\makebox[2.5cm]{{\sl Fermi\,}/GBM\cc} \\  \hline
&&&&\\ [-3mm]
Beginning of the event\dd, s & 0.44 &\multicolumn{2}{c|}{2.0}& 1.7 \\
Total duration\dd, s &6.0 &\multicolumn{2}{c|}{0.1}&4.1 \\
Integral number of counts&$2300\pm 420$&\multicolumn{2}{c|}{$570\pm120$}&--- \\ 
Significance (ratio $S/N$), $\sigma$ & 5.5      &\multicolumn{2}{c|}{4.6} & 8.7 \\ 
Probability\ee & $1.9\times10^{-8}$ &\multicolumn{2}{c|}{$2.1\times 10^{-6}$}&$1.7\times10^{-18}$\\ 
FAR\ff, events/s &$6.4\times 10^{-5}$&\multicolumn{2}{c|}{$4.2\times 10^{-4}$}&---\\
Combined probability\gg &$1.6\times10^{-4}$&\multicolumn{2}{c|}{$4.8\times10^{-3}$}&---\\
Fluence $F$\hh, erg cm$^{-2}$&$8.0\times 10^{-8}-2.4\times
10^{-6}$&\multicolumn{2}{c|}{$1.7\times
  10^{-8}-5.2\times10^{-7}$}&$ (2.1\pm 0.3)\times 10^{-7}$\\
Energy release $E_{\rm iso}$\ii, erg&$2.2\times 10^{47}-6.7\times 10^{48}$&
\multicolumn{2}{c|}{$3.8\times10^{45}-1.2\times10^{47}$}&$(4.7 \pm 0.7)\times 10^{46}$\\ \hline
\multicolumn{5}{l}{}\\ [-1mm]
\multicolumn{5}{l}{\aa\ The moment of event registration by
  LIGO/Virgo detectors, UTC.}\\ 
\multicolumn{5}{l}{\bb\ Area of the localization region for the
  event}\\
\multicolumn{5}{l}{\cc\ According to Goldstein et al. (2017), Pozanenko et al. (2018).}\\
\multicolumn{5}{l}{\dd\ The beginning (since the moment $T_ {0}$) and total
  duration of the gamma-ray burst.}\\
\multicolumn{4}{l}{\ee\ Probability for the pulse to be detected
  by chance assuming Gaussian statistics for $S/N$.}\\
\multicolumn{5}{l}{\ff\ False Alarm Rate for the random events of
  such temporal structure (according to data from the entire orbit).}\\   
\multicolumn{5}{l}{\gg\ Probability which takes into account
  identification with the gravitational-wave event by chance and}\\   
\multicolumn{5}{l}{\ \ \ enumerating time series with different bin sizes (Blackburn et al. 2015, $T_{\rm min}=0.1$ s, $T_{\rm max}=30$ s).}\\ 
\multicolumn{5}{l}{\hh\ Fluence in the 10--1000 keV band.}\\ 
\multicolumn{5}{l}{\ii\ Equivalent isotropic energy radiated during the burst.}\\ 
\end{tabular}
\end{table*}

\section*{THE S190425Z EVENT}
\noindent
The gravitational-wave event S190425z was recorded by the
LIGO/Virgo detectors on April 25, 2019 at 08\uh18\um05\fsec017
UTC. With a reliability of $>99$\%, it was assigned to the
events caused by the BNS merger (Singer 2019a), becoming the
second act of such a merger discovered in the entire history of
observations. The False Alarm Rate of such events (FAR) was
estimated very low FAR\,$ = 4.5\times 10^{- 13}\ \mbox{s}^{- 1}$
or 1 event per 69834 years\footnote{\sl gracedb.ligo.org/superevents/S190425z\/}.

At the time of registration, only two detectors of the
gravitational-wave interferometer were operating: LIGO L1
(Livingston, USA) and Virgo V1 (Italy). Accordingly,
localization of the event was much more uncertain than in the
case of the GW\,170817 event (Abbott et al. 2017a). Area of the
50\%-region of localization was 1378 sq. deg, while that of the
90\%-region was 7461 sq. deg (Singer 2019b). The region is
divided into two parts of nearly equal size --- north and south
(see below Fig.\,\ref{fig:locmap}). The source turned out to be
4 times more distant than GW\,170817, it was at a distance of
$156\pm41$ Mpc. This further complicated the search for its
manifestations in all ranges of the electromagnetic
spectrum. The characteristics of the S190425z event are given in
Table\,1. For comparison, there you can find similar data on the
GW\,170817 event.

Immediately after the announcement on the S190425z event and its
identification (Singer 2019a) optical, soft X-ray and radio
telescopes around the world were involved in searching for a
possible afterglow of this object or a kilonova which could
flare at the location of the BNS merge (e.g., Abbott et
al. 2017b). Some early results of this study were presented by
Coughlin et al. (2019) and Hosseinzadeh et al. (2019). Although
it is already obvious that the corresponding event was not
quickly detected in these energy ranges, there is no doubt that
these are only the first ones in the stream of works based on the
results of such studies.

\begin{figure}[t]
  \includegraphics[width=0.99\linewidth]{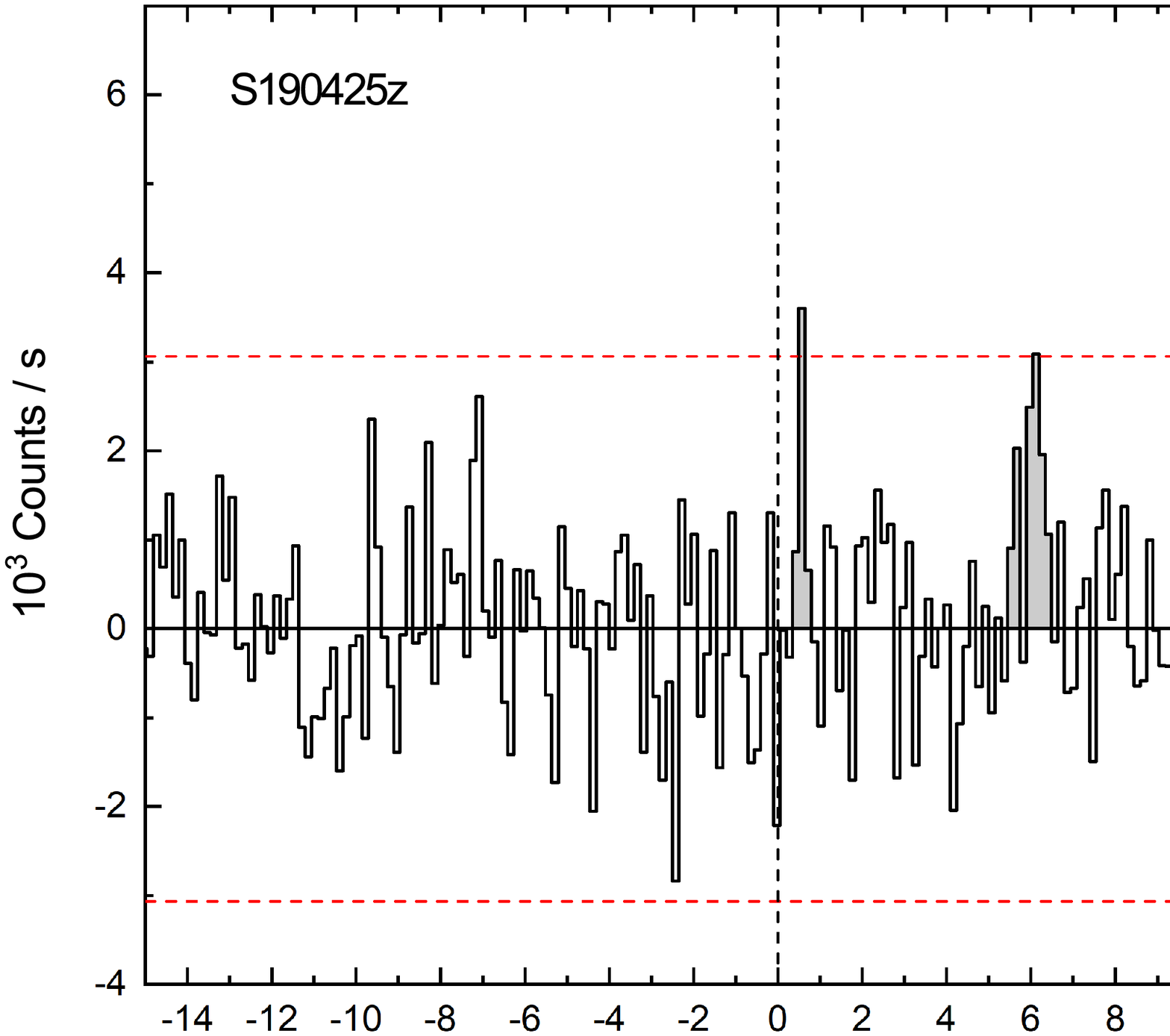}
  \includegraphics[width=0.99\linewidth]{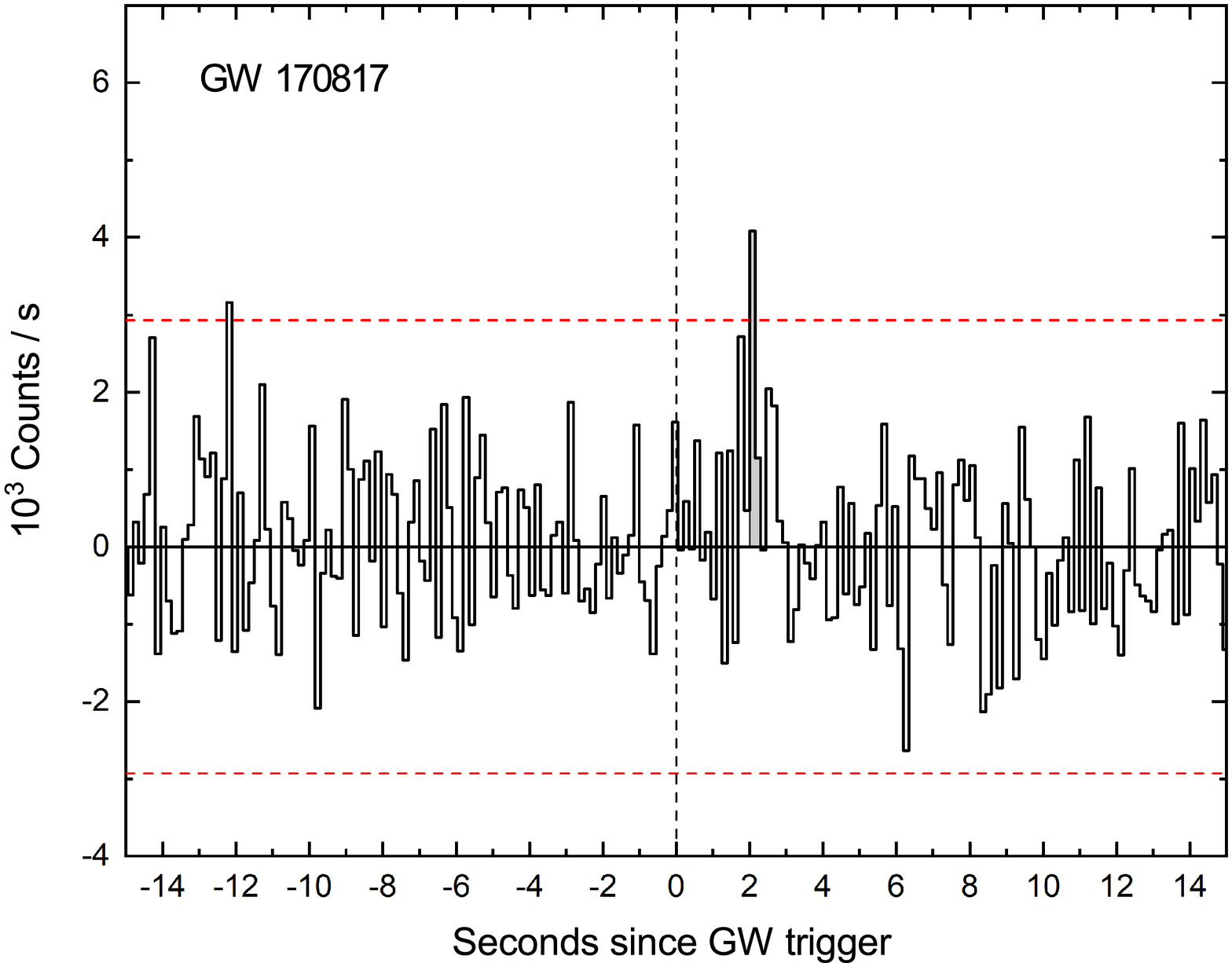}
\caption{\rm Photon count rate by SPI-ACS versus time
  immediately before and after the gravitational-wave events
  S190425z (top) and GW\,170817 (bottom), i.e. the same as in
  Fig.\,\protect\ref{fig:lcurve_0.85}, but in a narrower time 
  interval ($ \pm15 $ s) and with a bin duration 0.15~s. Time
  is counted from the trigger by the LIGO/Virgo detectors
  (indicated by the vertical dashed line), the horizontal dashed
  (red) lines indicate the level of $3\sigma$ for an accidental excess.
  \label{fig:lcurve_0.15}}

\vspace{-0.5cm}
\end{figure}

\section{RESULTS}
\noindent
As the top panel of Fig.\,\ref{fig:lcurve_0.85} shows and as it
was first reported by Martin-Carilla et al. (2019), Minaev et
al. (2019a), the SPI-ACS detector on board INTEGRAL recorded a
significant enchancement in the count rate history over the
background in $\sim5.94$~s after the moment $T_0$ of the
S190425z event (Singer 2019a). The bin size in the count rate
history in this figure is 0.85~s. A priori confidence of its
registration (the signal-to-noise ratio corrected for
non-Poissonity in count rate of the SPI-ACS detector) is $S /
N\simeq 4.4$ standard deviations (Table\,2). The figure clearly
shows that the excess is really significant --- in the time
interval of 500 s in duration with a center at $T_0 $ there is
no excesses even at $3 \sigma$ (the level shown by the
dashed red line) besides the named one.

The bottom panel of the figure shows a similar history of photon
count rate near the GRB\,170817A event. No significant excesses
of the count rate exist in it. This is due to the fact that the
duration of the gamma-ray burst detected in the SPI-ACS data from
this event was much smaller ($\simeq 0.1$~s, Savchenko et
al. 2017c; Pozanenko et al. 2018) than the selected bin size. If
we consider a record of the count rate with a smaller bin size ---
0.15~s, see the bottom panel of Fig.\,\ref{fig:lcurve_0.15}, a
significant ($S/N\simeq 4.3$) excess corresponding to
GRB\,170817A appears. When reducing the size of the bin to 0.1~s
its significance reaches a maximum of $S/N\simeq 4.6$
(Table\,1).

Amazingly, there is another significant (with $S/N\simeq 3.6$)
pulse in $\sim0.5$~s after the moment $T_0$ of the
gravitational-wave event S190425z present on the count rate
history recorded with the 0.15~s step shown in the top panel of
Fig.\,\ref{fig:lcurve_0.15} (Minaev et al. 2019). The total
duration of this pulse reaches $\sim 0.5$~s, although the
maximum of its radiation is contained in a very narrow ($\simeq
0.15$~s) peak. The significance of the second pulse on this
light curve noticeably decreased (to $S/N\simeq3.1$ in one
bin). This is not surprising with such a small devision of the
time series, because the actual duration of the second pulse
reaches $\sim1.3$~s.

Thus, there were two significant excesses over the background
level found (Fig.\,\ref{fig:lcurve_0.15}, top panel) in the
interval $\pm30$~s near $T_0$ for this event. Further, we
will call them the first and second pulses in the time
profile of this gamma-ray burst, which thus has a total 
duration of $\sim6.0$~s. The overall significance of the double
event is $S/N\simeq 5.5$ (see Table\,1).

The estimate of fluence from such a burst given in the table,
with an average value of $F_{\rm m}\simeq 4.4\times
10^{-7}\ \mbox{erg cm}^{-2}$ in the 10--1000 keV range was
obtained taking into account normalization of counts recorded by
the SPI-ACS detector in respect of the fluences measured by the
{\sl Fermi\,}/GBM monitor for a number of short gamma-ray
bursts, simultaneously detected by both the instruments (see
Attachment). It will be discussed in more detail below.

Each pulse reaches maximum significance on the light
curve with its well-defined size of bins: $0.15$~s for the first
and $0.85$~s for the second pulse. Both the pulses have an
equally high significance of $S/N\simeq3.5$ and $3.3$ standard
deviations on the light curve with $0.25$-s bins.
Joint probability taking into account both the statistical
significance of pulses and probability of their association with
the S190425z event by chance (as well as an increase in the
number of tests due to the selection of the optimal size of
bins), is computed below precisely for such a light curve. The
main parameters of both pulses of the burst are given in
Table\,2.

\begin{figure}[t]
\centerline{\includegraphics[width=0.99\linewidth]{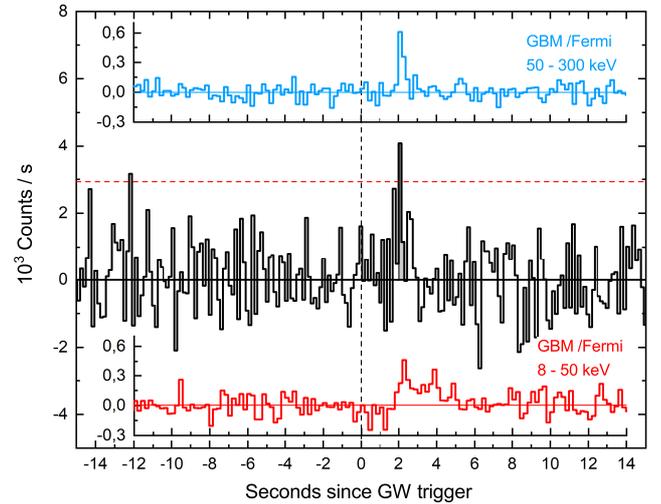}}
\caption{\rm Temporal profile of the gamma-ray burst
  GRB\,170817A, accompanying the gravitational-wave event
  GW\,170817, according to the SPI-ACS detector (black, $\ga80$
  keV, see the bottom panel in
  Fig.\,\protect\ref{fig:lcurve_0.15}) and {\sl Fermi\,}/GBM in
  the soft (red, 8--50 keV) and hard (blue, 50--300 keV)
  ranges. It is obvious that in the soft range the burst
  continued until at least $\sim6$~s after the merge of the
  neutron stars.
  \label{fig:lcurve_gbm}}
\end{figure}

For comparison, the similar parameters for the GRB\,170817A
burst according to the SPI-ACS detector are also given
there. According to the light curve in Fig.\,\ref{fig:lcurve_0.15}
this burst seems to contain a single rather narrow pulse. In fact,
as was noted in the work by Pozanenko et al. (2018), the SPI-ACS
detector has recorded only its initial hard part. According to
data of the {\sl Fermi\,}/GBM monitor (Goldstein et al. 2017)
GRB\,170817A had a much longer profile in the X-ray 8--50 keV
range with a total duration of $\sim4.1$ s. This is clearly seen
in Fig.\,\ref{fig:lcurve_gbm} where the light curves near this
event, recorded by the {\sl Fermi\,}/GBM (red and blue
histograms), are shown in comparison with the light curve
recorded by the SPI-ACS detector (black histogram).
In the range of 50--300 keV of this instrument, close to the
SPI-ACS range ($> 80$ keV), the narrow initial gamma-ray pulse
dominates in the burst profile, but there is also an indication
for the presence of a weak second pulse with a duration of $\sim
1$~s in $\sim 5.3$~s after the moment of merging the neutron stars.

So the gamma-ray bursts that accompanied both nearby
gravitational-wave events recorded to the moment have 
comparable durations and the same two-component structure of the 
time profile.

Note that there is no any extended emission that could be
regarded as the afterglow up to 250~s after the moment $T_0 $ in
the light curve recorded by the SPI-ACS detector (see
Fig.\,\ref{fig:lcurve_0.85}). There was also no extended
emission recorded in the light curve of GRB\,170817A (Pozanenko
et al. 2018). There is no gamma-ray emission, that could be
associated with the gravitational-wave event, recorded in the
SPI telescope itself. However, the registration of the emission
by the SPI-ACS shield implies that gamma-ray photons come at a
large angle to the telescope's axis and could not be recorded
within its field of view.

\subsection*{Evaluation of the event's confidence}
\noindent
To evaluate the probability that two pulses on the hard X-ray
light curve immediately after the moment $T_0$ of the S190425z
event's registration appeared by chance, we used the
two-parameter formula (Blackburn et al. 2015, 2019), which takes
into account both the statistical confidence of the pulses and
the probability of their random association with the event
S190425z, i.e. their appearance in a certain time interval after
the event. The formula also takes into account the enhancement
of the probability of finding a significant random pulse due to
the increase of the number of trials resulting from our
selection of the optimal time step in the count rate
recording. For the first time this formula was applied to the
evaluation of the significance of the weak transient gamma-ray
burst recorded by the {\sl Fermi\,}/GBM monitor shortly after
the first LIGO/Virgo gravitational-wave event GW\,150914
(Connaughton et al.  2016).

The evaluation is carried out in several stages. First, the
False Alarm Rate (FAR) is computed --- the empirically
determined rate of occurrence of random events (pulses) on the
light curve with significance equal to or exceeding a certain
value. Then, the probability for such an event to occur no later
than the time ${\rm d}\,T$ after $T_0$ is computed. The estimate
of probability is not directly dependent on the step (bin) of
the light curve at which the events (pulses) were searched for.
Let us evaluate the rate of occurrence of a complex of two
pulses with the parameters corresponding to the gamma-ray burst
GRB\,190425 (Tables\,1 and 2). For this we use the SPI-ACS data
obtained during the entire orbit of the INTEGRAL observatory at
which the burst was recorded (rev.\,2083). We will investigate
the count rate record with 0.25~s steps (the time series
consists of 591200 bins). In total during 125~ks there were 8
such complexes found in the record with a distance between the
beginnings of two pulses to be less than 5.5~s. The short pulse
preceded the longer one in 5 complexes and lagged it in the rest
of the cases. Thus,
$$\mbox{FAR}\simeq \frac{8}{1.25\times 10^5\ \mbox{\rm s}} = 6.4\times
10^{-5}\ \mbox{s}^{-1}.$$

Joint, very conservative (overstated) estimate of the
probability of coincidental in time occurrence of the random
pulse (Blackburn et al. 2015; Connaughton et al. 2016) is written as
$$
P = \mbox{FAR}\  \times \ln (1+T_{\rm max}/T_{\rm min})\,\mbox{\rm d}\,T, 
$$ where $\mbox{\rm d}\,T$ is the duration of a time interval
since $T_0$ till the beginning of the first pulse; $T_{\rm max}$
is the duration of interval of the time series after $T_0$ at
which the events were searched for; $T_{\rm min}$ is
the duration of the minimum measurable match. Conservatively,
it is possible to limit $T_{\rm max}$ to 30~s; $T_{\rm min}$ is
obviously the minimum bin of the time series for which
pulses were searched for, which is 0.1~s  in our study (see also
Connaughton et al. 2016). Substituting these values, 
we obtain the following estimate of the probability 
$$ P = 6.4\times 10^{-5}\ \ln \left(1+
\frac{30}{0.1}\right)\times 0.44\simeq 1.6\times 10^{-4}.
$$ This estimate reflects the probability of the accidental
occurrence of a complex of two pulses of a given intensity at a
distance of 0.44~s after the gravitational-wave event S190425z.

The similar estimate can be obtained separately for each of the
recorded pulses. The results are given in Table\,2.  Note that
there were 198 positive and 139 negative excesses found above
the level of $S/N=3.5$ on the light curve with a bin size of
0.25~s (591\,200 bins in total, the curve for the entire
revolution). The number of negative excesses is in a good
agreement with that expected for Gaussian statistics with $P (>
3.5 \sigma) \simeq 2.3 \times10^{-4}$, the number of positive
ones exceeds it by 40\%.

Most likely this is due to the presence of a noticeable number
of pulses of high intensity, associated with charged particles,
in the SPI-ACS count rate.  They give only positive
excesses. In this case, excesses of low confidence ($S/N\la3$)
occur to be consistent with Gaussian probability. It is obvious
that under such conditions only the empirical estimates should
be used in order to determine the probability of recording a
random excess (see Table\,2).

There were 4 positive (including the second pulse of
GRB\,190425) and 1 negative excesses having a signal to noise
level $S/N = 4.4$ or exceeding it found on the light curve with
a bin size of 0.85 s (containing only 173\,900 bins). For
Gaussian statistics with the probability $P (> 4.4 \sigma)
\simeq 5.4 \times10^{-6}$ we would register only 1 random
excess for 125~ks.

\subsection{Comparison with the confidence of GRB\,170817A}
\noindent
For comparison with the estimate of the GRB\,190425 confidence
let us carry out a similar analysis of data for the first
gamma-ray burst GRB\,170817A associated with the GW\,170817
event of the BNS merger recorded by the SPI-ACS detector. Let us
recall that the maximum significance of recording this burst was
$S/N\simeq 4.6$ on the light curve with a bin size of 0.1~s
(Table\,1).

In the record of the detector's count rate with such bin size
obtained during the whole revolution of the INTEGRAL observatory
corresponding to this gamma-ray burst (rev.\,1851) which had a
duration of 155~ks (1\,549\,000 bins) there were 65 positive
pulses recorded with $S/N\geq$4.6 and 3 negative ones. With
Gaussian statistics, the probability to record by chance such a
pulse is equal to $2.1 \times 10^{-6}$, i.e. we had to record 3
positive pulses during 155~ks.  The false pulses connected with
charged particles dramatically worsen the statistics. According to
the measured number of false pulses for this event FAR$ =
4.2\times10^{-4}\ \mbox{s}^{-1}$ (Table\,2).

A conservative estimate for the probability of the random
coincidence for this event, carried out by the method of
two-parameter analysis (Blackburn et al. 2015; Connaughton et
al. 2016), gives $$P = 4.2 \times 10^{-4}\ \ln \left
(1+\frac{30}{0.1}\right) \times 2.0 \simeq4.8 \times10^{-3}.$$
The estimate does not exclude random origin of the
event. However, the reliability of the gamma-ray burst
GRB\,170817A has been confirmed independently --- by its simultaneous
recording by the {\sl Fermi\,}/GBM monitor with much higher
significance of $S/N \simeq 8.7$ (Table\,1).

\begin{figure}[t] 
\centerline{\includegraphics[width=0.99\linewidth]{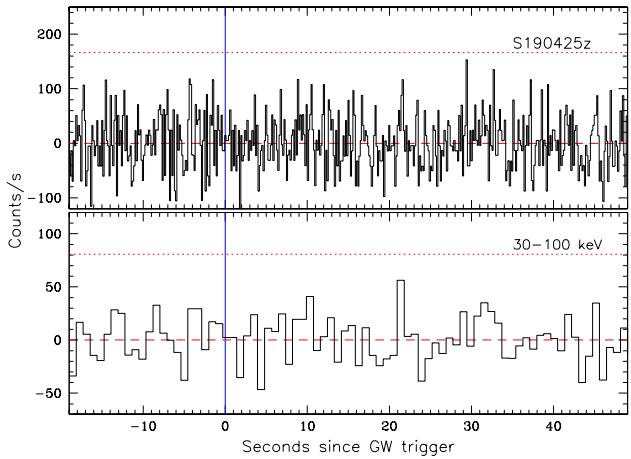}}
\caption{\rm Photon count rate by the IBIS-ISGRI detector versus
  time since $-20$~s till $+50$~s after the
  gravitational-wave event S190425z. The bin duration is
  equal to 0.15~s (top panel) and 0.85~s (bottom panel), time is
  counted since the trigger times of the LIGO/Virgo detectors,
  the background was subtracted,  the red dotted line marks the
  range of random deviations at the level of $3\sigma$.
\label{fig:lcurve_ibis}}
\end{figure}

\subsection{Observations by the IBIS-ISGRI telescope}
\noindent
Figure\,\ref{fig:lcurve_ibis} shows the time dependence of the
photon count rate recorded by the ISGRI detector of the IBIS
telescope aboard the INTEGRAL observatory near the event
S190425z. The top and bottom panels of the figure correspond to
different choice of the time step length for these curves, 0.15
and 0.85 s, respectively. The data were taken in the 30--100 keV
energy range. It is obvious that there were no significant
bursts of radiation found in the count rate record since
$T_0-20$~s till $T_0+50$~s that could be interpreted as a
continuation of the gamma-ray burst GRB\,190425 in the hard
X-ray range. There were no bursts of radiation found on a longer
time scale either, see Chelovekov et al. (2019a), Savchenko et
al. (2019).

Of course, we could not expect the gamma-ray burst, recorded by
the SPI-ACS detector, to be detected in the field of view of the
telescope. However, as was recently shown by Chelovekov et
al. (2019b), the IBIS-ISGRI telescope is able to successfully
detect the bursts coming from the side, at large angles to its
axis, therefore, some emission from GRB\,190425 could have been
detected. The upper limit ($3\sigma$) for the flux of any
possible excessive emission in the 10--1000 keV range lasting
$\sim1$~s is $2.1\times10^{-6}\ \mbox{erg cm}^{-2}$ (if using
data of the IBIS-ISGRI detector in the energy range 30--100 keV)
and $1.2\times 10^{-6}\ \mbox{erg cm}^{-2}$ (if using data in
the range 100--500 keV). To estimate the flux from the burst
presumably coming at an angle of 26\deg\ -- 60\deg\ to the
telescope axis, we used the normalization based on the analysis
of several hundred gamma-ray bursts observed simultaneously by
the {\sl Fermi\,}/GBM and IBIS-ISGRI detectors (Chelovekov et
al. 2019b). The obtained limit does not contradict to the flux
measured from the burst by SPI-ACS.

\subsection*{Observations of S190425z by other experiments}
\noindent
In none of the other X-ray and gamma-ray experiments, including
SWIFT/BAT (Sakamoto et al. 2019), MAXI/GSC (Sugizaki et
al. 2019), WIND/KONUS (Svinkin et al. 2019), AGILE/MCAL
(Casentini et al. 2019), {\sl Fermi\,}/GBM (Fletcher 2019),
Insight-HXMT/HE (Xiao et al. 2019), the burst GRB\,190425 from
the LIGO/Virgo event S190425z has been detected. However,
for all these instruments, except {\sl Fermi\,}/GBM, the flux
recorded by SPI-ACS (see Table\,1), was obviously below the
detection threshold. The obtained $3\sigma$ upper limits for the
flux of pulsed emission with a duration of 1\,s were at best
comparable to the limits set by the IBIS-ISGRI telescope and
more often have notably exceeded it.

According to our analysis of a representative sample of the
short bursts detected simultaneously by the {\sl Fermi\,}/GBM
and INTEGRAL/SPI-ACS detectors (see Appendix and
Fig.\,\ref{fig:gbm_calib}), a gamma-ray burst with
characteristics of the event under consideration should have been
definetely detected by {\sl Fermi\,}/GBM.  Nevertheless, this
instrument has not detected the burst within the interval of
$\pm30$~s from the moment $T_0$ of the gravitational-wave event,
the $3\sigma$ limit on its fluence in the 10--1000 keV range
was, depending on the used spectral model, $(0.9-8.4)\times
10^{-7}\ \mbox{erg cm}^{-2}$ for the very short (duration
$\sim0.1$ s) burst, $(0.3-2.5)\times 10^{-6}\ \mbox{erg
  cm}^{-2}$ for the typical short ($\sim1$ s) burst, and
$(0.9-7.7) \times10^{-6}\ \mbox{erg cm}^{-2}$ for the long
($\sim10$ s) burst (Fletcher 2019).

\begin{figure*}[t] 
\centerline{\includegraphics[width=0.96\linewidth]{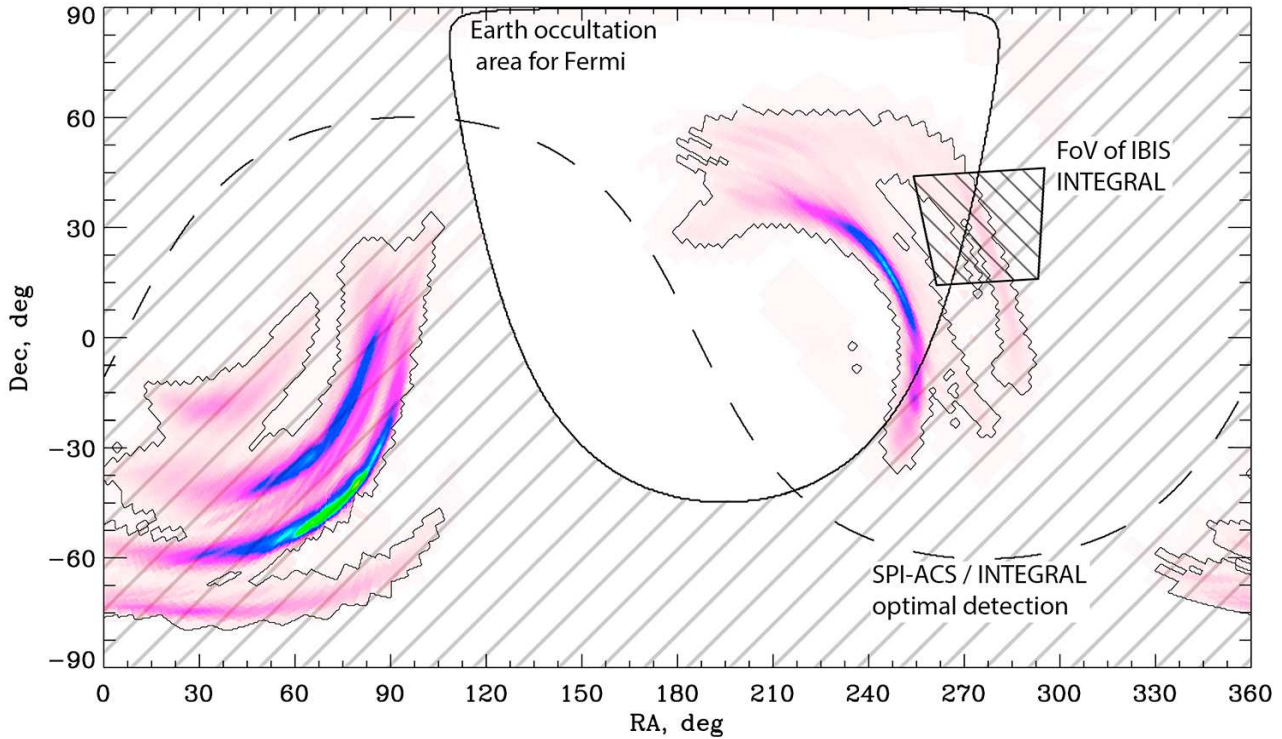}}
\caption{\rm Map of regions for the most probable localization
  of the gravitational-wave event S190425z by the LIGO/Virgo
  detectors ({\sl gracedb.ligo.org/superevents/S190425z/view},
  blue and green correspond to the maximum probability).  The
  map shows the area of occultaion of the field of view of the
  {\sl Fermi\,}/GBM monitor by the Earth at UTC 2019-04-25
  08:18:05, and also the area falling at this time in the field
  of view of the IBIS-ISGRI telescope of the INTEGRAL
  observatory. The source of the gamma-ray burst can be located
  only in the unshaded part of the map --- in the northern
  region of its localization by LIGO/Virgo. The dashed line
  indicates the strip for optimal recording the event by the
  SPI-ACS detector of the INTEGRAL observatory (the shown are
  the central points the strip which are located at an angle of
  90\deg\ to the axis of the telescope).
\label{fig:locmap}}
\end{figure*}

\section*{DISCUSSION}

We believe that the described excess in the count rate of the SPI-ACS
detector immediately after the gravitational-wave event
S190425z of merging the neutron stars was indeed associated with
registration of a gamma-ray burst from its source. In this
regard it is appropriate to note the following observational facts.

\subsection*{The lack of detection by {\sl Fermi\,}/GBM}
\noindent
The lack of detection of the gamma-ray burst GRB\,190425 by the
{\sl Fermi\,}/GBM monitor can be explained by the occultation of
its source with the Earth. According to Fletcher (2019) the {\sl
  Fermi\,}/GBM observations have covered only 56\% of the region
of the initial localization of the source by the LIGO/Virgo
detectors at the time of S190425z.

Fig.\,\ref{fig:locmap} shows an updated map of the localization
region for the S190425z event with the LIGO/Virgo detectors
(Singer 2019b) and the region of the occultation by the Earth of
the \mbox{\sl Fermi\,} spacecraft (unshaded area). Indeed, it is
obvious that almost all northern part of the localization region
of the gravitational wave signal was occulted by the Earth at the
moment of the gamma-ray burst registration.  The curve of
optimal detection of the burst with SPI-ACS (at an angle of
90\deg\ to the telescope's axis) is shown by a dashed line. The
detector is sensitive to the events in a wide band, at least as
distant as $\pm75$\deg\ from this curve; the entire zone of
maximum probability from the northern region for the S190425z
localization falls in this band.

\begin{table*}[t]
\noindent
{\bf Table 2.} Comparison of parameters of the gamma-ray bursts,
associated with neutron star merger events GW\,170817 and
S190425z, according to the SPI-ACS experiment\\ [3mm]
\begin{center}
\begin{tabular}{l|c|c|c}\hline\hline
  &\multicolumn{2}{c|}{}&\\ [-3mm]
\multicolumn{1}{c|}{Event}&\multicolumn{2}{c|}{GRB\,190425}&\multicolumn{1}{c}{GRB\,170817A}\\ \hline
&&&\\ [-3mm]
Pulse on the event profile   &first&second                &first\\
Beginning of the pulse $T_{i}$\aa, s  & 0.44              &5.54 & 2.00 \\ 
Time of the maximum count rate $T_{\rm m}$\aa, s  & 0.54              &5.94 & 2.05 \\
Binning $N_{i}$\bb                          & 5                    & 17   &2 \\
Duration $\Delta T_{i}$\cc, s           &0.25               &0.85 &0.10\\
Integral number of counts $C_{i}$ in the pulse $i$ \ &$700\pm200$ &$1600\pm370$&$570\pm120$\\
Significance (ratio $S/N$), $\sigma$ & 3.5                 & 4.4 & 4.6 \\ 
Probability\dd     &$2.3\times10^{-4}$& $5.4\times10^{-6}$& $2.1\times10^{-6}$\\
FAR\ee, events/s &$1.4\times10^{-3}$&$2.7\times10^{-5}$&$4.2\times10^{-4}$\\
Conservative probability\ff&$3.5\times10^{-3}$&$8.5\times10^{-4}$&$4.8\times10^{-3}$\\
\hline \multicolumn{4}{l}{}\\ [-1mm]
\multicolumn{4}{l}{\aa\ The beginning of the pulse $i$ and its
  maximum (from the moment $T_0$ of the GW event).}\\
\multicolumn{4}{l}{\bb\ Optimal binning of the initial time series
  having $50$-ms bins.}\\ 
\multicolumn{4}{l}{\cc\ The corresponding optimal duration of bins
    $\Delta T_{i}=0.05\ N_{i}$ s}\\
\multicolumn{4}{l}{\ \ \  (characterizes the duration of the
  given pulse).}\\ 
\multicolumn{4}{l}{\dd\ Probability to be detected by chance
  according to Gaussian
  statistics for $S/N$.}\\
\multicolumn{4}{l}{\ee\ False Alarm Rate for such random events
  (according to data from the entire orbit).}\\
\multicolumn{4}{l}{\ff\ Probability which takes into account
  association with the GW event by chance and enumerating}\\ 
\multicolumn{4}{l}{\ \ \  time series with different
  bin sizes (Blackburn et al. 2015, $T_{\rm min}=0.1$ s, $T_{\rm max}=30$ s).}\\ 
\end{tabular}
\end{center}
\end{table*}
Thus, the intersection of the {\sl Fermi\,} region occulting by
the Earth with the localization region of the event by the
LIGO/Virgo detectors is the area where the optical source
(afterglow of the gamma-ray burst or a kilonova) possibly
accompanying the S190425z neutron star merger is likely
located. We can also exclude the region in the field of view of
the IBIS-ISGRI telescope in which the burst would be necessarily
detected (the shaded trapezoidal region centered at R.A.$ \simeq
277$\deg, Decl.$ \simeq 30$\deg). The field of view of the SPI
spectrometer practically coincides with the field of view of
IBIS-ISGRI.

\subsection{Similarities and differences of GRB\,190425 and GRB\,170817A}
\noindent
Characteristics of the gamma-ray bursts GRB\,170817A and
GRB\,190425 detected in the SPI-ACS experiment are given in
comparison in Tables 1 and 2.

The GRB\,190425 and GRB\,170817A gamma-ray bursts are similar in
that they both consisted of two episodes --- pulses: the first,
short (in case GRB\,170817A only it was detected by SPI-ACS),
and second, longer. In the case of GRB\,170817A the second pulse
lasted almost 4~s, in the case of GRB\,190425 --- $1.3$~s (with
the maximum in 5.4~s after the first pulse, see Table 2). The
total duration of both the gamma-ray bursts was comparable and
equal to $\sim 4-6$~s. Extended pulses have been observed in the
time profile of a number of other short gamma-ray bursts
(Gehrels et al. 2006 and references therein), effectively
increasing their duration.

The second pulse of GRB\,170817A was noticeably softer than the
first (see Fig.\,\ref{fig:lcurve_gbm} and in more detail ---
Pozanenko et al. 2018). The second pulse of GRB\,190425 remained
hard enough, that is why it was detected by SPI-ACS, which has a
lower energy threshold of sensitivity as high as $\sim80$
keV. If this pulse were as soft as the second pulse of
GRB\,170817A ($kT\sim 11$ keV) then SPI-ACS would not have been
able to detect it.

Moreover, as was shown by Gottlieb et al. (2018), Pozanenko et
al. (2018), the second pulse in the GRB\,170817A temporal
profile was, most likely, associated with the thermal heating of
the shell at the jet breakout (with the cocoon's radiation). Since
the distance to the GRB\,190425 source ($\sim 156$ Mpc)
significantly exceeds the distance to the GRB\,170817А source
($\sim 40$ Mpc, see Table 1), the intensity of the thermal
component in the emission spectrum of GRB\,190425 should have
been in $(156/40)^2\sim 15$ times lower than the intensity of
the thermal component in the GRB\,170817A spectrum, so it could
not have been registered not only with SPI-ACS, but even with
{\sl Fermi\,}/GBM. It is quite possible that the second pulses
in these two bursts had a different origin.

Given that in the case of GRB\,190425 two pulses are separated
from each other by 5.4~s, a two-jet gamma-ray burst scenario
could be implemented in this case (Barkov and Pozanenko 2011), where
the first short pulse corresponds to the jet resulting from the
neutrino annihilation (Chen and Beloborodov 2007) and the
second, more prolonged one, appears as a result of accretion from
the formed accretion disk and the Blanford-Znajek (1977)
effect. In this case, the angle at which an observer sees the
jet of the gamma-ray burst 190425, should be smaller than the angle
of observation of the jet in GRB\,170817A. Thus, the nature of
the two observed episodes of radiation in the light curves of
GRB\,170817A and GRB\,190425 may be different. 

\begin{figure*}[t] 
  \centerline{\includegraphics[width=0.96\linewidth]{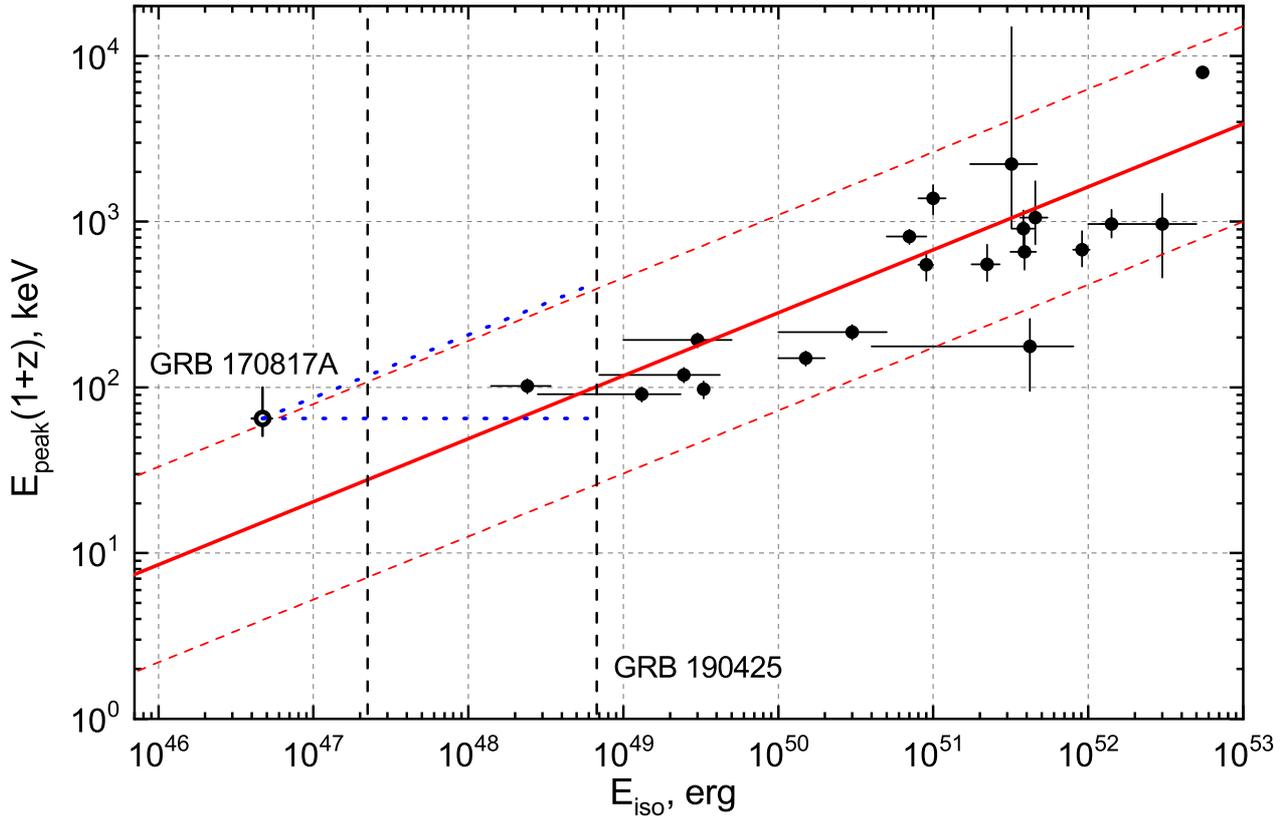}}
\caption{\rm Possible position of the GRB\,190425 burst,
  accompanying the event S190425z of the neutron star merge
  recorded by the LIGO/Virgo detectors, on the Amati (2002)
  diagram, constructed taking into account only the short bursts
  (Minaev and Pozanenko 2019). The vertical dashed lines indicate the
  boundaries of the $2\sigma$-region of uncertainty of the
  $E_{\rm iso}$ value for the GRB\,190425 burst. They were
  obtained from the results of SPI-ACS calibration (see
  Appendix) by recalculating the fluence $F$ into $E_{\rm iso}$
  for a photometric distance of 156 Mpc ($E_{\rm
    iso,min}=2.2\times 10^{47}$ erg and $E_{\rm
    iso,max}=6.7\times10^{48}$ erg). The intersection of these
  lines with the $2\sigma$-region of uncertainty of the $E_{\rm
    p}-E_{\rm iso}$ dependence gives the maximum possible values
  of $E_{\rm p}$ (at the level of $2\sigma$): $E_{\rm p,min}=7$
  and $E_{\rm p,max}=400$ keV. The blue dotted lines
  additionally restrict $E_{\rm p}$ under assumption that the
  energy emitted during the bursts GRB\,170817A and GRB\,190425
  was the same, and all differences in their observed
  manifestations are associated only with different orientations
  of axes of their relativistic ejections (jets) in respect to
  the observer.
\label{fig:amati}}
\end{figure*}

\subsection*{Classification and spectral properties}
\noindent
There is no doubt that both the gamma-ray bursts belong to the
group of type I bursts (also called short bursts), having
merging neutron stars as their sources and progenitors. This
follows from the observations and data analysis of LIGO/Virgo.

Although SPI-ACS does not have spectral channels, some
conclusions about spectral properties of the gamma-ray emission
of GRB\,190425 can be given. So, using the empirical dependence
``the energy of the maximum $E_{\rm p}$ in the energy spectrum
of $\nu\,F_{\nu}$ --- the equivalent isotropic radiated energy
in the gamma-ray range $E_{\rm iso}$'' (Amati 2002) for Type I
gamma-ray bursts (Minaev and Pozanenko 2019) the limits can be
obtained within which the energy $E_{\rm p}(1+z)$ for
GRB\,190425 must lie (Fig.\,\ref{fig:amati}). The solid line in
this figure shows the best dependence
$$E_{\rm p}(1+z)\simeq105\ \left(\frac{E_{\rm
    iso}}{10^{49}\ \mbox{erg}}\right)^{0.38\pm0.06}\ \mbox{keV},$$
while the dashed (red) lines show the $\pm2\sigma$ area of scattering
the real bursts relatively this dependence. The redshift of the
burst source is $z\simeq 0.0364 \ll 1,$ therefore the factor $(1
+ z)$ will be further neglected.

The energy $E_{\rm iso}$ is determined based on estimates of the
minimum/maximum possible fluence from the burst reduced to the
10--1000 keV {\sl Fermi\,}/GBM energy range. The estimates were
computed from the fluence measured by SPI-ACS (in counts, see
Table\,1), using the results of calibration of the ratio of
fluences from a number of short gamma-ray bursts measured by the
{\sl Fermi\,}/GBM monitor and simultaneously by the SPI-ACS
detector (see Appendix and Fig.\,\ref{fig:gbm_calib}).

The boundary values of the fluence were then transformed to
$E_{\rm iso}$ taking into account the 156 Mpc photometric
distance to the source. The obtained limits of $E_{\rm iso, min} =
2.2 \times 10^{47}$ erg and $E_{\rm iso, max} = 6.7 \times
10^{48}$ erg are shown in Fig.\,\ref{fig:amati} with the
vertical dashed lines. Their intersection with the
$2\sigma$-region of spreading the bursts relatively to the
$E_{\rm p}-E_{\rm iso}$ dependence gives the maximum possible
$E_{\rm p}$ values at the $2\sigma$ confidence level: $E_{\rm p,
  min} = 7$ and $E_{\rm p, max} = 400$ keV (correspond to the
lower and upper corner of the zone of intersection of the
uncertainty strips in Fig.\,\ref{fig:amati}).

Assuming that the dependence $E_{\rm p}-E_{\rm iso}$ is
associated with the geometry of observations of the source of
the gamma-ray burst, and namely, with the value of the angle
between the axis of the relativistic ejection (jet) and
direction to the observer (Eichler and Levinson 2004; Levinson  
and Eichler 2005; Ito et al. 2015, 2019), the additional
restrictions on the value of $E_{\rm p}$ can be obtained. 

Indeed, suppose that the total energy emitted in the gamma-ray
range during the GRB\,170817A and GRB\,190425 bursts was
approximately the same. Then the angle $\theta$ between the jet
axis and direction to the observer in the source of GRB\,190425
should be less than this angle in the source of GRB\,170817A
(see Song et al. 2019), since $E_{\rm iso}$ for GRB\,190425 is
many times larger than $E_{\rm iso}$ for GRB\,170817A (see
Fig.\,\ref{fig:amati}).

The decrease of possible values of the angle $\theta$ with
increasing $E_{\rm iso}$ is confirmed by detailed calculations
in frameworks of the jet model with Gaussian profile (Zhang,
Mezaros 2002; Troy et al. 2018). Conservatively $E_{\rm p}$ can
be restricted from the bottom by the value $E_{\rm p,min} = 70$
keV, because it is necessary for the presence of overall
positive correlation $E_{\rm p} \sim E_{\rm iso}^{\alpha}$,
where \mbox{$\alpha>0$} (the lower dashed blue line in the
figure). The upper limit $E_{\rm p,max} = 400$ keV will remain
the same, it is restricted by the uncertainty of the observed
relationship $E_{\rm p}-E_{\rm iso}$ (the upper dashed
line). The obtained limits do not contradict the GRB\,190425
registration by the SPI-ACS detector at energies above 80 keV.

\section*{CONCLUSION}
\noindent
The report by Minaev et al. (2019) on the detection of a
possible gamma-ray burst in the time interval of 0.5--6.0 s
after the S190425z gravitational-wave event by the SPI-ACS
detector of the INTEGRAL observatory remained almost unnoticed.
The absence of registration of this burst by the {\sl
  Fermi\,}/GBM monitor could be responsible for this. The other
possible reason could lie in a not serious enough assessment of
reality of the burst by Martin-Carillo et al. (2019) and
Savchenko et al. (2019), based on their analysis of the same
SPI-ACS data.  In this work we confirm the sufficiently high
statistical confidence of the burst, explain the lack of its
registration by the {\sl Fermi\,}/GBM and list a number of
additional arguments supporting its actuality.

\begin{enumerate}
\item The GRB\,190425 gamma-ray burst was recorded by the
  SPI-ACS detector 0.44~s after the detection of the S190425z
  gravitational-wave event. The burst consisted of two emission
  pulses (episodes) lasting 0.25~s and 0.85~s (the second pulse
  started 5.1~s after the first one). The burst had the total
  duration ($\sim6.0$~s) and time profile, in many aspects
  similar to the duration and profile of the GRB\,170817A
  gamma-ray burst accompanying GW\,170817 --- the first event
  recorded by LIGO/Virgo from the BNS merger.

\item The joint probability of an accidental appearance of the
  complex, consisting of two pulses described above, is $1.6
  \times 10^{-4}$. This probability takes into account along
  with the usual significance $S/N \simeq 5.5 \sigma$ of the
  double burst the possibility of its false association with
  the S190425z event and an increase in the number of trials
  due to our selection of an optimal time scale (Blackburn et
  al. 2015). For comparison, the probability of recording the
  GRB\,170817A burst with a duration of 0.1~s by the SPI-ACS detector
  with the significance $S/N \simeq 4.6\sigma $ in $\sim2.0$~s
  after the GW\,170817 event is $4.8\times 10^{-3}$.

\item Both sources of the detected bursts, GRB\,170817A and
  GRB\,190425, are at the distances (40 and 156 Mpc,
  respectively) smaller than the distances to other events
  of the BNS mergers (as well as BBH and NSBH mergers) recorded
  in the O2 and O3 observing cycles of LIGO and LIGO/Virgo.
  
\item There were no significant evidences of the presence
  of gamma-ray radiation in any of the individual events of BBH
  or NSBH mergers registered by the LIGO/Virgo detectors (Savchenko
  et al. 2016, 2017b, 2018).

\item The obtained conservative estimate of the isotropic energy
  $E_{\rm iso}$ emitted during GRB\,190425 is bounded by the
  $2\sigma$ range, from $2.2 \times 10^{47}$ to $6.8 \times
  10^{48}$ erg, which is at least 5 times higher than the estimate of
  $E_{\rm iso}$ for GRB\,170817A. The estimate for the energy
  $E_{\rm p}$ of the maximum in the GRB\,190425 emission
  spectrum is bounded by the $2 \sigma$ range from 70 to 400 keV
  (Fig.\,\ref{fig:amati}).

\item Since $E_{\rm iso}$ for GRB\,190425 exceeds $E_{\rm iso}$
  for GRB\,170817A significantly, the angle between the
  direction to an observer and the jet axis in GRB\,190425z
  should be smaller than that in GRB\,170817A (e.g. Song et
  al. 2019). This is obvious under the assumption of the same
  energy radiated during the bursts and is confirmed by
  computations in frameworks of the model of Gaussian profile of
  the jet (Zhang, Meszaros 2002; Troja et al. 2018).  The
  estimate of the angle to the jet axis is an independent
  estimate of the angle between the direction to an observer and
  the orbital plane of the binary system of merging neutron
  stars (under the assumption that the jet axis is perpendicular
  to the orbital plane of the system). This angle is poorly
  defined directly from the gravitational-wave observations.
  
\item The absence of registration of the  GRB\,190425 gamma-ray burst
  by the {\sl Fermi\,}/GBM monitor (one of the most
  sensitive omnidirectional gamma-ray burst experiments in the
  range above 10 keV) can be explained by the fact that its
  source was occulted by the Earth at the moment of the burst.

\item As a result of overlapping the localization area of the
  S190425z event with the LIGO/Virgo detectors and the area
  shaded for {\sl Fermi\,}/GBM by the Earth during GRB\,190425,
  the region of possible localization of the event is
  significantly reduced (in comparison with the original
  localization area by LIGO/Virgo) and consists only of its
  northern part (Fig.\,\ref{fig:locmap}).

\item The lack of registration of an optical component of the
  event in the form of afterglow, apart from the large size of
  the localization region (7461 sq. deg.), may be associated
  with noticeable deviation of the jet axis from the direction
  to an observer, which leads to exponential suppression of the
  flux being recorded. The large distance to the source
  ($\sim156$ Mpc) does not allow the possible volume of its
  localization to be effectively studied. Targeted observations
  of galaxies in this volume are complicated by incompleteness
  of existing catalogs of galaxies, while the survey
  observations do not provide sensitivity necessary for
  registration of the transient.  So, intense follow-up
  observations aimed at finding the kilonova (e.g. the ZTF and
  Palomar Gattini-FR telescopes have covered $\sim20$\% of the
  localization region of the event, Coughlin et al. 2019, and
  the MMT and SOAR telescopes --- 40\% of the possible volume of
  localization, Hosseinzadeh et al. 2019), were unsuccessful,
  perhaps for particularly these reasons.

  To continue the search for an optical companion of the 
  S190425z event it in necessary to concentrate efforts on the
   area of its localization in the north hemisphere refined in
   this work and examine more closely optical transients
   discovered in it. 
\end{enumerate}
\begin{figure*}[t]
\centerline{\includegraphics[width=0.96\linewidth]{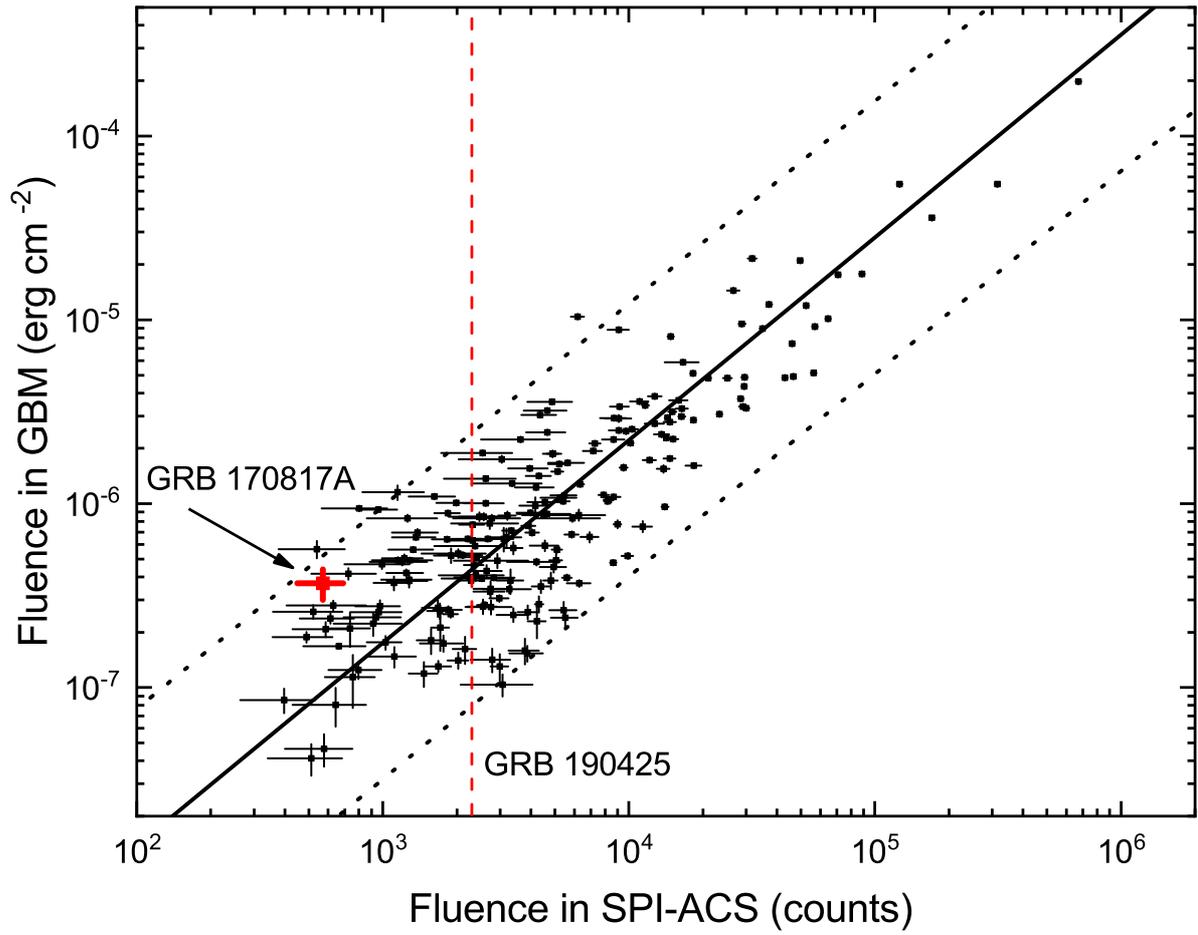}}
\caption{\rm Relation between the time-integrated photon flux
  (in counts) recorded by the SPI-ACS detector of the INTEGRAL
  observatory in the range $>80$ keV from the short ($T_{90} <6$
  s) gamma-ray bursts, and the time-integrated radiation flux
  (fluence) measured by the {\sl Fermi\,}/GBM monitor from the
  same bursts in the 10--1000 keV range (in erg cm$^{-2}$).  The
  dashed lines bound the $\pm2\sigma$ region of deviations of
  the fluence measured by the {\sl Fermi\,}/GBM monitor from the
  best-fit approximations of this ratio (shown by a solid
  line). The intersection of these dashed lines with the
  vertical (red) dashed line sets the range of possible values
  of the 10--1000 keV fluence from GRB\,190425 (associated with
  the gravitational-wave event S190425z).
\label{fig:gbm_calib}}
\end{figure*}

\vspace{3mm}

\section*{ACKNOWLEDGMENTS}
\noindent
This work is based on the observations performed by
the \mbox{INTEGRAL} international gamma-ray astrophysics
observatory and retrieved via its Russian and European
Scientific Data Centers. We are grateful to the Russian Science
Foundation for financial support (grant no. 18-12-00522).

\begin{appendix}
\section{FLUX CALIBRATION FROM THE SHORT BURSTS IN THE SPI-ACS
  DETECTOR} 
\noindent
To estimate the fluence corresponding to the integrated number
of counts recorded by the SPI-ACS detector during GRB\,190425,
we selected and investigated a representative sample of short
gamma-ray bursts observed simultaneously with the
INTEGRAL/SPI-ACS detector and the {\sl Fermi\,}/GBM monitor. As
a result, the relation has been established between the fluences
measured by these instruments (in counts and erg~cm$^{- 2}$,
respectively).

There were the events selected for the sample from the burst
catalog (Bhat et al. 2016) of the {\sl Fermi\,}/GBM
monitor\footnote{The persistenty updated catalog of bursts of
  this monitor can be found at the www-address {\sl
    heasarc.gsfc.nasa.gov/W3Browse/fermi/fermigbrst.html\/}}
recorded since July 14, 2008 till June 30, 2019 and
contained at least one outlier at the $>3\sigma$ level of
significance over the average count rate in the interval
$T_{90}$\footnote {Time interval between the moments of
  accumulation of 5\% and 95\% of the integrated number of
  counts by the GBM monitor (Koshut et al. 1996).} on the
SPI-ACS light curve obtained with a 50-ms step. Taking into
account the origin of the GRB\,170817A and GRB\,190425 bursts
discussed in the paper and their actual measured durations, only
the bursts with $T_{90}<6$ s as determined by the GBM monitor
were used for the sample.

For all 278 bursts of the sample selected in such a way, the
background count rate of photons by the SPI-ACS detector was
approximated by a polynomial of the 3rd degree in the intervals
($T_0-300$~s, $ T_0-50$ s) and ($T_0+200$ s, $ T_0 + 500 $ s)
--- separately in each interval; the start time of the burst
$T_0 $ was taken from the catalog of the GBM monitor. The
average value of two model count rates at the boundaries of the
interval ($ T_0-50 $ s, $ T_0 + 200 $ s) was used as the
background count rate $B$ in this interval. The integrated
number of counts $C$ recorded during the burst has been computed
as the total excess of the count rate over $B$ in the time
interval $T_{\rm 100}$ \footnote {The interval used for building
  the spectrum and calculating the fluence for a given burst in
  the catalog of the GBM monitor.}. Only bursts with $C$
determined with the significance over 3 standard deviations were
chosen for further analysis

In Fig.\,\ref{fig:gbm_calib} the fluences $F$ (in erg cm$^{-2}$)
recorded by the {\sl Fermi\,}/GBM monitor during the chosen
bursts are given as a function of the integrated number of
counts $C$, recorded from these bursts by the SPI-ACS
detector. This dependence can be approximated by the power law
model function
$$
F_{\rm m}= 2.19\times10^{-6}
\left(\frac{C}{10^4\ \mbox{counts}}\right)^{1.10\pm0.06}\ \mbox{erg
  cm}^{-2},
$$ shown in Fig.\,\ref{fig:gbm_calib} by the solid line. The
dashed lines show the $\pm2\sigma$ deviation area of the
fluences actually measured from the gamma-ray bursts relative
to this line. Previously, a similar dependence has been obtained
by Vigano and Mereghetti (2009), but for a more limited (by the
number of bursts) and less uniform (taking into account both short
and long bursts) sample.

The vertical red dashed line shows the possible position of
GRB\,190425, corresponding to the event S190425z, on this
dependence, according to the integrated number of counts
measured by the SPI-ACS detector. The intersection of this line
with two dashed lines bordering the strip of uncertainty of the
model dependence $F_{\rm m}(C)$ sets the range (at confidence
level of $2\sigma$) for the fluence from this burst in the
10--1000 keV energy range: $F_{\rm min}\simeq 8.0\times
10^{-8}\ \mbox{erg cm}^{-2}$, $F_{\rm max}\simeq 2.4\times
10^{-6}\ \mbox{erg cm}^{-2}$ (see Table\,1).

The red cross in this figure shows the position of the gamma-ray
burst GRB\,170817A, corresponding to the event GW\,170817, and
the errors of measurement of its fluence by the two
instruments. The strong shift of the burst position to the left
(and to the up) relative to the line of best approximation of
the dependence $ F_{\rm m}(C)$ probably reflects the fact of
incomplete registration of photons from the burst by the SPI-ACS
detector due to the already noted softness of its radiation
(Pozanenko et al. 2018; recall that SPI-ACS is sensitive above
80 keV).
\end{appendix}

\vspace{1cm} 


\begin{flushright}
{\sl Translated by authors\/}
\end{flushright}

\end{document}